\documentclass[aps,onecolumn,nofootinbib]{revtex4-2}
\usepackage{amssymb,amsfonts,amsmath}
\usepackage{lipsum}
\usepackage[usenames]{xcolor}
\usepackage{adjustbox}
\usepackage{graphicx}
\usepackage{microtype}
\usepackage{epsfig}
\usepackage{setspace}
\usepackage{layout}
\usepackage{textcomp}
\usepackage{hyperref}
\usepackage{amsfonts}
\usepackage[T1]{fontenc}
\usepackage{lmodern}

\usepackage{multirow}
\usepackage[usenames]{xcolor}
\usepackage[lined,boxed,commentsnumbered]{algorithm2e}
\usepackage{amsthm}
\usepackage{lineno}

\makeatletter
\providecommand*{\cupdot}{%
  \mathbin{%
    \mathpalette\@cupdot{}%
  }%
}
\newcommand*{\@cupdot}[2]{%
  \ooalign{%
    $\m@th#1\cup$\cr
    \hidewidth$\m@th#1\cdot$\hidewidth
  }%
}
\makeatother

\newtheorem{theorem}{Theorem}

\newtheorem{proposition}{Proposition}
\newtheorem{definition}{Definition}

\renewenvironment{proof}{{\bf \emph{Proof of} }}{\hfill $\Box$ \\}

\newcommand{\R}		{\mathbb{R}}
\newcommand{\N}		{\mathbb{N}}

\renewcommand{\Pr}[1]	{\mathbb{P}\left( #1 \right)}
\newcommand{\Ex}[1]	{\mathbb{E}\left( #1 \right)}
\newcommand{\Var}[1]	{\mathbb{V}ar\left( #1 \right)}

\begin{document}
 \bibliographystyle{plain}

\title{A self-organized criticality participative pricing mechanism for selling zero-marginal cost products}


\author{Daniel Fraiman$^{1,2}$}
\affiliation{$^1$ Departamento de Matem\'atica y Ciencias, Universidad de San Andr\'es, Buenos Aires, Argentina,}
\affiliation{$^2$ CONICET, Argentina.}
\email{dfraiman@udesa.edu.ar}


\begin{abstract}
In today's economy, selling a new zero-marginal cost product is a real challenge, as it is difficult to determine a product's ``correct'' sales price based on its profit and dissemination. As an example, think of the price of a new app or video game. New sales mechanisms for selling this type of product need to be designed, in particular ones that consider consumer preferences and reality. Current auction mechanisms establish a time deadline for the auction to take place. This deadline is set to increase the number of bidders and thus the final offering price. Consumers want to obtain the product as quickly as possible from the moment they become interested in it, and this time does not always coincide with the seller's deadline. Naturally, consumers also want to pay a price they consider ``fair''. Here we introduce an auction model where buyers continuously place bids and the challenge is to decide quickly whether or not to accept them. The model does not include a deadline for placing bids, and exhibits self-organized criticality; it presents a critical price from which a bid is accepted with probability one, and avalanches of sales above this value are observed. This model is especially interesting for startup companies interested in profit as well as making the product known on the market.    \\ \newline

\end{abstract}
\maketitle
\section{Introduction}

How to determine the price of a zero-marginal product is still an open question for marketing and economics. 
An auction~\cite{auction} is an efficient mechanism for establishing this value. In an auction, the ``true'' value of a product is unknown to the seller, and the sales price is discovered during the course of open competitive bidding. The two most well-known auction mechanisms are the Blind auction and the English auction. In the former, bidders place their bid in a sealed envelope and simultaneously hand them to the auctioneer. The envelopes are opened and the individual with the highest bid wins, paying the amount bid. On the other hand, the English auction is an open-outcry ascending dynamic auction.

In this ascending pricing auction, the auctioneer calls out a low price for a \textit{single or multi-item} product and raises it until there is only one interested buyer remaining. Traditionally used to sell rare collectibles and antiques, nowadays auctions are widely used with many types of products and services, both in the traditional way (live) and on the Web (e.g. Amazon and eBay). For example, Google, Microsoft, and Yahoo! use auctions for search advertising, and Amazon uses an auction mechanism for selling computing time on the Cloud. Despite these apparent superficial differences, most auction mechanisms share a common core: they are designed to sell products or services within a restricted time framework~\cite{gallego}. For example, when selling a publicity spot set for Monday 07/26/2021 between 20:00h and 20:01h, the purchase must be made before the product ``expires''. As a more common example, the seller establishes an arbitrary deadline time for selling a product, bidders compete, and the final decision is made at the deadline time. Much has been written about the difference between the various auction mechanisms and which is best (highest purchase price)~\cite{sub1,sub2,sub3,sub4}. However, there is much less research concerning auction mechanisms designed for a real situation where people are interested in a product at different moments of time that are not controlled by the seller. For example, think of our everyday consumption pattern. Most of what we consume is not regulated by some deadline established by a seller. We only purchase when we are aware of the existence of the product, decide we ``need'' it, and are able to buy it.  While sellers sometimes use promotions and discounts that may encourage consumers to decide to purchase, this effect is different from that of a true deadline time established at an auction. Here, we propose an auction mechanism that considers the possibility that buyers may be interested in buying the product at any time.

The auction model presented here is particularly interesting for companies that sell products with a zero-marginal cost. An important part of today's economy is based on doing business with products or goods that have zero or almost zero marginal cost. Examples of these products include adding a student to an online course, selling one more mobile application, adding a Facebook account, offering a new home on Airbnb, responding to a new customer via a virtual agent, having a car in UBER, and, in the future, analyzing a medical image using an algorithm. Establishing the price of these products is not trivial, and perhaps one of the most important tasks is making the product known on the market. Importantly, developers of zero-marginal cost products develop the product looking to sell a very large number of the product. That is why here we focus on the asymptotic regime of the model.

Let us suppose that a company places on the market a new useful mobile app (i.e., a zero marginal cost and infinite stock product). The selling price may be very low or zero but it may be difficult for customers to find it in a ``sea of applications''. Besides earning a profit, the company that sells this product is interested in increasing its visibility in the marketplace and earning market share when it encounters competitors. Here we present an auction model for selling these products in a time-continuous way, considering the customers' arrival times and bids. The model assumes that demand can be easily met (as happens with software when stock is unlimited), and that there is a certain degree of hurry in making the decision if the bid is accepted. The model belongs to a class of auctions sometimes referred to as digital goods auctions~\cite{goldberg,digital}.

\section{Literature Review}

Innovations in price management are mainly driven by novel products or services that aim to be launched in the market. New business models are continuously developed to sell these products or services. In the last 15 years, several sales techniques have been developed for selling products or services with zero-marginal-cost or near to zero marginal cost. Among the most well-known mechanisms, we can mention: \textit{Flat Rate}, \textit{Freemium},  \textit{Name your own price}, and \textit{Pay what you want}.

In the flat rate sale method, customers pay a fixed price per occasion or time period and can use the acquired service or goods or as much as they choose. Today, this sale technique is used in music, video streaming, internet and mobile telephone services, and is also a standard sales method for sport clubs and for some museums as part of a special membership. This sale technique faces a big problem when the percentage of heavy users starts to grow. Moreover, the flat rate price regulates how much will be consumed by the customers. Basically, customers are motivated to get their money's worth. This last effect has been shown in an experiment where customers went to an All-you-can-eat restaurant~\cite{Flatrate1}.

The freemium is a widely used pricing strategy for online products. Customers can use a basic version of a product or service for free or pay a fee to use a premium version of the same product or service. Today this pricing strategy is mainly used for selling apps, games, media, entertainment, social networks, and software. This sale mechanism has two main positive aspects: (a) it allows the customer to have a first idea of what the product is about with no money down, which (b) results in a large market penetration. However, the premium version (the one that generates revenues), typically does not have a large market penetration ~\cite{Freemium1,Freemium2,Freemium3,Freemium4,Freemium5,Freemium6,Freemium7,Freemium8} (i.e., only a small fraction of the free version customers decide to switch to premium). Since the conversion rate from Free to Premium is typically very small, many companies try to make some money by selling advertisements in the free version.

In the Name-Your-Own-Price (NYOP) sales technique, the customer offers a price and the seller decides whether to accept it. In order to make the sales decision, the seller fixes a threshold price for the product or service -- let us call it $P$ -- which is unknown by the customer. The customer makes an offer $X$, and if $X$ is greater than $P$, then he/she buys the product and the seller gets an ``extra'' $X-P$ revenue.

The most famous website to start with this sales technique is Priceline.com. Jay Walker founded this company in 1997 and was motivated by the question: ``What if leisure travelers could use the Internet to find last-minute deals offered by hotels and airlines that want to clear excess inventory?''. The problem was that the website's consumers were mostly bargain hunters and/or consumers that do not reveal their true willingness to pay in an attempt to acquire products at extremely low prices.

In the ``Pay what you want'' (PWYW) selling method, buyers pay their desired amount for the product or service on sale, which is frequently zero. In some cases, a minimum price may be set, or a suggested price may be indicated to guide the buyer. However, in this method, the seller no longer has any decision power. This selling method was put on the spot by the band RadioHead, who sold their album \textit{In Rainbows} in this way. Since then, several articles~\cite{Pay8,Pay7,Pay3,Pay1,Pay2,Pay4,Pay5,Pay6} have analyzed the pros and cons of this method, trying to elucidate which variables affect the price offered by the customer.

Among all variables, the most important ones are the buyer-seller sympathy/commitment relationship~\cite{Pay8,Pay7,Pay3}, whether there is a reference price for the same or similar product~\cite{Pay3,Pay8}, and face-to-face interactions between buyer and seller~\cite{Pay8}. The PWYW method is obviously a risky selling strategy, but depending on the specific product or service being sold (and/or the company), the proportion of consumers that act fairly toward the company may be large~\cite{Pay5,Pay6}, and can even lead to an increase in sales revenues. This selling method has been applied successfully in some zoos, restaurants (clients decide onthe prices for their meals), hotels, and software products.

Interestingly, in~\cite{Pay9}, the authors found that participatory pricing methods led to a greater intent to purchase (i.e., more customers are interested in buying the product). Therefore, new participative pricing mechanisms can produce a larger total income (even if the average price paid for the product is smaller than the price set by the seller in a non-participative mechanism).

The sales method proposed in this paper can be considered to be somewhere between the \textit{Name your own price} method and \textit{Pay what you want}. In both methods, the interested buyers make an offer, but in the first, the seller controls the price, and in the latter, the buyer does. In the method proposed here the price is controlled by the whole community of interested buyers (not by one individual buyer). The whole community, by competitive behavior, determines a threshold price, a distinctive feature of the auction mechanism presented here. Unlike known auctions, in this case it is not the seller who sets the threshold price. This threshold value is unknown to buyers, as occurs in the \textit{Name your own price} method. Finally, as in the \textit{Name your own price} method, a bidder that offers a value greater than the threshold will get the product.

\section{The sales mechanism}

Now, let us suppose that the seller sells a product that has no production limitation (i.e., the seller can produce at the rate of demand, or stock is unlimited), such as digital products (a downloadable software program or mobile application). Next, let us suppose that each interested buyer can only make one price offer as in a Blind auction, and that the decision to purchase must be made with some degree of hurry.

Bidders appear at different times, which are described by some general stochastic process (e.g., an inhomogeneous Poisson Process with rate $\lambda(t)$, or any other). Once a bidder appears, he/she offers a bid price for the product. This value cannot be subsequently modified, and the bidder cannot participate again in the buy-sell process. The bidder does not know the bids made by the previous interested buyers, as is the case in a Blind auction. The seller will sell the product to the bidders with the largest bids, and the transaction decision must occur almost in real time. The precise mechanism is described below.

Let us suppose that the selling process starts at time zero and buyers start to appear. The first potential buyer (who appears at some arbitrary time $t_1$) offers a price $X_1$ for the product; the second potential buyer offers $X_2$, and so on.

\textit{The selling rule is the following: at each bid appearance time, the highest remaining bid is executed, except when the new bid exceeds this value. Whenever this happens, no transaction occurs and the new bid remains in the bid queue until it becomes the highest value. }

For example, let us suppose that the following bid values were offered in this order:
$$\$14,  \framebox[1.1\width] {\$15}_{ \textbf{3}}, \framebox[1.1\width] {\$18}_{  \textbf{1}}, \$13,   \framebox[1.1\width] {\$16}_{  \textbf{2}},  \$12 ,  \$10. $$
In this example, the first transaction occurs when the fourth person (bid) appears. When he/she offers a price of \$13 that is lower than the maximum at that time (\$18), the third bid value of \$18 is executed. The fifth bid value is the greatest value among the pending offers, which is why no transaction occurs at that moment. However, this last bid is executed during the following bid appearance time, when a smaller bid value of \$12 is presented. Finally, the bid value of \$15 is executed because it turns out to be the highest offer at the time the last bidder appears. As we saw in this example, only the three largest offers were executed. This sales process continues (up to infinity) because new buyers appear continuously, and the main goal is to understand how many products will be sold and how much money the company will earn when $N$ bidders make their offers. In particular, we are interested in understanding what the stationary regime of this process is.

\subsection{Mathematical description}
Let $X_1,X_2,X_3,\dots$ be the sequence of bid prices in the order they appear, and let $$\mathbb{X}_k=\{X_1,X_2,\dots, X_k\},$$ be the set that contains the first $k$ bid prices. The dynamics of the sales mechanism is described as follows.

The number of accepted bids, $\tilde{N}$, when $k+1$ bids have been offered,
\begin{equation}\label{BSS}
	\tilde{N}(k+1)=\left\{
	\begin{array}{lll}
		\tilde{N}(k)+1 &  &   \mbox{if}\ \ X_{k+1} < X^{max}_k \\
      \tilde{N}(k) &  & \mbox{if}     \ \ X_{k+1} \geq X^{max}_k \\
	\end{array}
	\right.
\end{equation}
where
\begin{equation}\label{BSSa}
X^{max}_k=max\{\mathbb{X}_k \setminus  \mathbb{Y}_k \},
\end{equation}
and
\begin{equation}\label{BSSb}
 \mathbb{Y}_k=\mathbb{Y}_{k-1} \cup \Delta_k,
\end{equation}
with
\begin{equation}\label{BSSc}
	\Delta_k=\left\{
	\begin{array}{lll}
		X^{max}_{k-1} &  &   \mbox{if}\ \ X_{k} < X^{max}_{k-1}\\ 	
\emptyset &  &   \mbox{if}\ \ X_{k} \geq X^{max}_{k-1}\\
	\end{array}
	\right.
\end{equation}
and initial condition
\begin{equation}\label{BSSd}
\mathbb{Y}_1=\emptyset \quad \mbox{and} \quad \tilde{N}(1)=0.
\end{equation}
The set $\mathbb{Y}_k$ contains all the prices of the accepted bids when $k$ bids have been offered. Note that
\begin{equation}
\tilde{N}(k)=|\mathbb{Y}_k|,
\end{equation}
and the Total Income, $TI$, when $k$ bids have been offered is
\begin{equation}
TI(k)=\underset{Y \in \mathbb{Y}_k}{\sum} Y.
\end{equation}
Note that eq.~\ref{BSS} does not have any parameter.

Now we present some definitions that will be useful for the following section.
The set of bids accepted when $k$ bids have been offered has already been defined as $\mathbb{Y}_k$.

\begin{definition}
Let $\mathbb{X}_k$ be the set that contains the first $k$ bids,
\begin{equation}
\mathbb{X}_k=\{X_1,X_2,\dots,X_k\}=\mathbb{Y}^{act}_k \ \cupdot \  \mathbb{Y}_k \ \cupdot \ \mathbb{X}^{frozen}_k,
\end{equation}
with the ``active'' set
\begin{equation}
\mathbb{Y}^{act}_k=\mathbb{X}_k \cap \mathbb{Y}_{k\to\infty},
\end{equation}
and the ``frozen'' set
\begin{equation}
\mathbb{X}^{frozen}_k=\mathbb{X}_k \setminus (\mathbb{Y}^{act}_k \cup \mathbb{Y}_k).
\end{equation}
\end{definition}
A bid $X_j$ that belongs to the frozen set $\mathbb{X}^{frozen}_{k}$ with $k\geq j$ is a bid that will never be accepted. A bid $X_j$ that belongs to the set $\mathbb{Y}_{k}$ with $k\geq j$, is a bid that has been accepted before the appearance of the (k+1)-th bid, and a bid $X_j$ that belongs to the active set $\mathbb{Y}^{act}_{k}$ with $k\geq j$, is bid that will be accepted in the future (i.e. there exists $k_1>k$ such that $X_j \in \mathbb{Y}_{k_2}$ for $k_2\geq k_1$).

Naturally, as $k$ grows the sets $\mathbb{Y}_{k}$ and $\mathbb{X}^{frozen}_{k}$ increases their sizes,
\begin{align*}
  & \underset{k\to\infty} {\lim} \Ex{|\mathbb{Y}_{k}|}=\infty, \\
  & \underset{k\to\infty} {\lim} \Ex{|\mathbb{X}^{frozen}_{k}|}=\infty.
\end{align*}
But the behavior of the size of $|\mathbb{Y}^{act}_{k}|$ is not obvious. We will show that although new bids become part of $\mathbb{Y}^{act}_{k}$ as $k$ grows, others converts into sales ($\in \mathbb{Y}_{k_1}$ with $k_1>k$). Namely, $\mathbb{Y}^{act}_{k}$ wins elements (bids) but it also loses elements that remain permanently in $\mathbb{Y}_{k_1}$.

\begin{definition}
Let define $Y_k$ as a randomly selected accepted bid from $\mathbb{Y}_{k}$. And let $Z_k$ be a randomly selected non-accepted bid from $\mathbb{X}^{frozen}_{k}\cup \mathbb{Y}^{act}_{k}$.
\end{definition}

\subsection{Bidder's behavior hypothesis}
As in a Blind auction, the interested buyer makes a unique offer based on his/her valuation of the product, which depends on multiple factors. Each bidder makes his/her own valuation without knowing either the previous bids or the bidders. Then, each participant offers a price valuation, $X$, that is well described by a certain probability density function $f(x)$ with cumulative probability $F(x)$. And since hypothetically bidders do not interact, then we can assert that the offered bids represent a sequence of independent random variables $X_1$, $X_2$, $X_3$, $\dots, X_n, \dots$ with probability law $F$. The key now is to understand the consequences of the new selling rule described above over this bid sequence.

\subsection{Questions about the model}
The following questions are studied in this paper: What is the stationary regime of the model? What is the expected number of sales and the expected Total Income? How does this model compare with some known sales models? What is the optimal bidder's strategy? What can be said about bidder collusion or bidders' strategic behavior? What type of modifications can be made to the model to preserve model dynamics?

The paper is organized as follows: in the next section: (1) we show the \textit{critical} behavior of the model, (2) we present a rigorous proof for the value $p_s$ and for the asymptotic  distributions of $Y_k$, $Z_k$ and $X^{max}_k$, (3) we study the asymptotic of the total income, (4) we discuss the sales process in terms of the Self-organized criticality theory (SOC) showing the existence of avalanches expected in SOC, (5) we compare our model with three other sales mechanisms, (6) and we discuss the valuation versus bid price problem and the robustness against the strategic behavior of bidders. Finally, in the last section, we summarize the results and present some considerations for the model's implementation.

\section{Results}
\subsection{Dynamics of the model and critical value}

A realization of the model (eq.1-5) is shown in Fig. 1 (A). This figure shows 1,000 bids, in the order they appeared, for the sale process with a Log-Normal price valuation distribution ($F(x)$). The red filled circles are the accepted bids and the black circles are the remaining ones.

 The right price histogram enhances the distribution of accepted versus not accepted bids.  Note that there is a sharp cut between the prices of accepted versus not accepted bids.  The cutoff appears at a particular value, called here critical price $x_c$ (represented by a dashed line). This important value verifies
\begin{equation}\label{xc}
	F(x_c)=p_c=:1-p_s,
\end{equation}
where $p_c$ ($p_s$) is the proportion of non-accepted (accepted) bids, and as we will see in Theorem 2 it verifies 
\begin{equation}\label{pc}
	p_c=1/e.
\end{equation}
The value $p_c$ is the asymptotic proportion of bidders that will not get the product (because they offered a low price). These non-successful bidders offer a value smaller than the critical price ($x_c$). Note that different price distributions ($F_X(x)$) give rise to different critical prices (eq.~\ref{xc}), but the dynamics remain unchanged. If we simulate the auction process for an exponential bid price distribution, $F(x)=1-e^{-\lambda x}$ (for $x>0$), the same dynamical behavior will be observed but now with a critical price equal $x_c=-\frac{1}{\lambda}\ln(1-p_c)$ and the same proportion $p_c$ of not accepted bids will be observed. This is because the dynamics of the model considers the order of the offers and not the exact values. We emphasize that the critical value $p_c$ is not a parameter of the model, it emerges from the model's own selling rule. The critical price ($x_c$) is not a parameter either, it depends on the price valuation distribution and $p_c$ through eq.~\ref{xc}.

\begin{figure}
	\begin{center} 
		\includegraphics[height=14cm,angle=0]{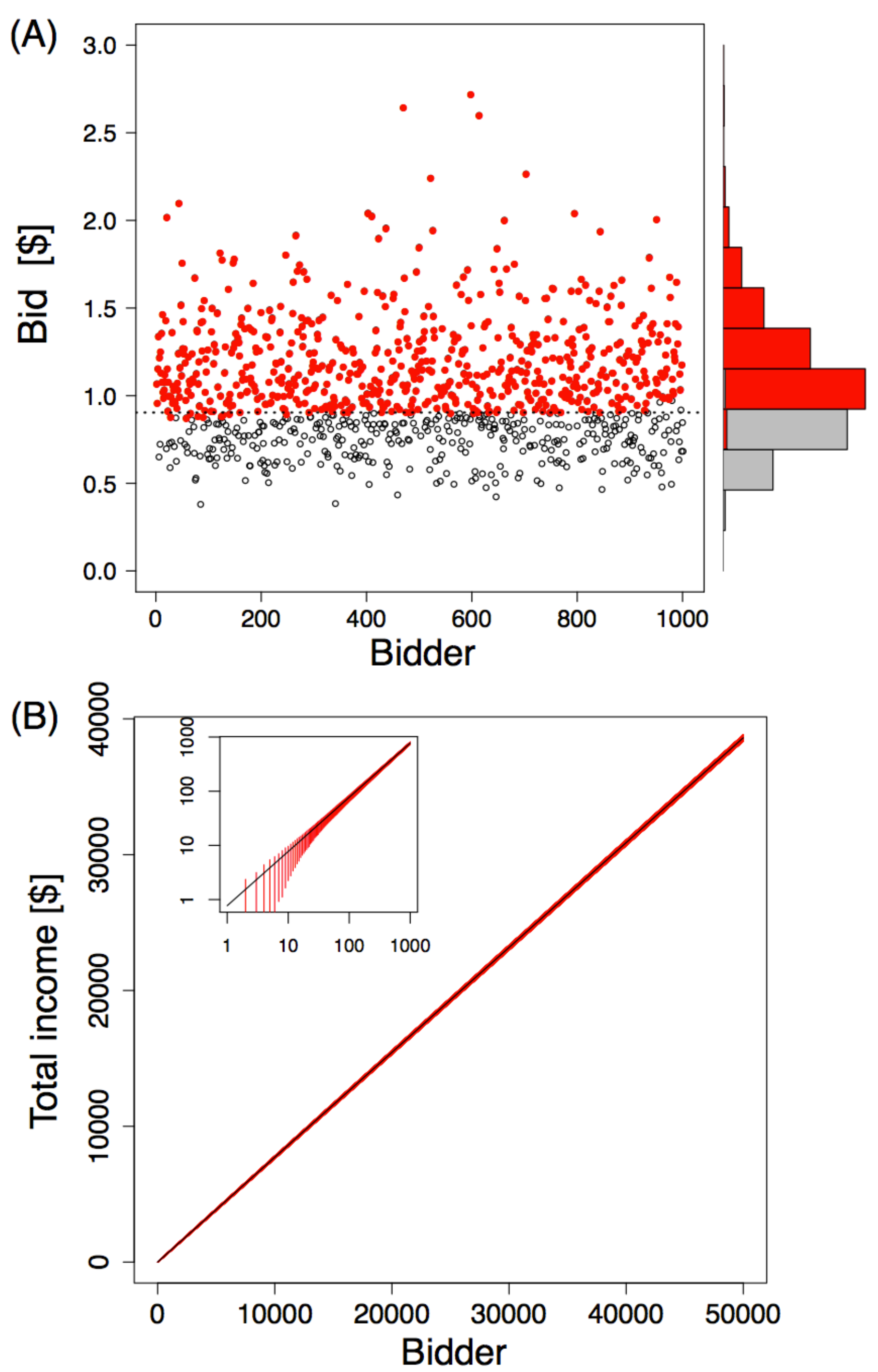}
	\caption{(A) The auction process for a Log-Normal price distribution, $f(x)=\frac{1}{x\sigma\sqrt{2\pi}}e^{-\frac{(ln(x)-\mu)^2}{2\sigma^2}}$,  with $\mu=0$ and $\sigma=0.3$ is shown. The offered prices (bids) are shown for 1,000 bidders in the order they appear. The accepted bids are shown with red filled circles, and the remaining bids with black circles.  The critical price $x_c$ ($F_X(x_c)=p_c=e^{-1}$ and therefore $x_c\approx 0.90371$) is shown with a dashed line.  A histogram of the bid values is shown on the right.  (B) An empirical 95\% confidence interval is shown for the Total Income (TI) in red. Two-hundred simulations were made and the interval \textit{average $\pm$ 3 standard deviation} is shown. The theoretical value $\Ex{TI}$ is shown with a black line ($\Ex{TI}=\int_{x_c}^{\infty}yf(y)dyN=0.7720651N$). In the inset we show a close-up of the first values together with the theoretical value. }
\end{center}
\end{figure}

Numerical simulations show that after a short transient period (or after a small number of bids) the system reaches equilibrium, where bids are accepted if they are greater than $x_c$. This behavior arises because the auction process leads to a situation where most of the remaining offers become ``frozen'' (they belongs to $\mathbb{X}^{frozen}_k$) and are never executed because the offered prices are low, and only a very small number of the remaining offers (belonging to $\mathbb{Y}^{act}_k$), will be executed in the near future. The surprising part of this model is that it presents a truly sharp separation between ``frozen'' and ``active'' bids. With probability one, bids greater than $x_c$ are accepted, and for large $N$, additionally, a small number of bids with a value just below $x_c$ can be accepted.

So far we have shown numerically that at some time long after the start of the auction, let us say when $N >> 1$ interested buyers have made their bids, an average of $(1-p_c)N$ bidders will have made a purchase
and each one has paid a value greater than a value $x_c$ which verifies eq.~\ref{xc}. Next, we study the critical value.

\begin{definition}
The expected proportion of sales when $n$ bidders made their offer is
\begin{equation}\label{pn}
p(n):=\frac{1}{n}\Ex{\tilde{N}(n)}.
\end{equation}
\end{definition}

\begin{definition}
The probability of having a sale when the $n-th$ bidder appears is
\begin{equation}\label{Pn}
P(n):=\Pr{X_n < X^{max}_{n-1}}.
\end{equation}
\end{definition}

Since the process reaches a stationary regime, according to the Ergodic Theorem, we know that the asymptotic proportion of sales verify
\begin{equation}\label{lim}
p_s=\underset{n\to \infty}{\lim} p(n)=\underset{n\to \infty}{\lim} P(n).
\end{equation}
From equations~\ref{BSS},\ref{BSSa},\ref{BSSb},\ref{BSSc},\ref{BSSd},\ref{pn} and~\ref{lim},  the asymptotic proportion of sales, $p_s$, can be obtained in principle, and also the asymptotic variance ($\underset{n\to \infty}{\lim}\frac{\Var{\tilde{N}(n)} }{n}$). Unfortunately, however, both limits are hard to solve, since the stochastic process has long-memory dependence.

The calculation of $P(n)$ can be seen as a combinatorial problem. For example, to calculate $P(3)$ we compute how many permutations of the sequence 1,2,3 (or any sequence $a_1,a_2, a_3$ of unequal values) has the last value of the sequence smaller than the maximum remaining value in second place. For $n=3$ this last value always corresponds to the second value ($X^{max}_2=X_2$). We will consider the sequences as vectors of dimension 3. In this case, there are only three permutations from a total of six (3!) that present this property. The sequences are (3,2,1), (2,3,1), and (1,3,2). Therefore, $P(3)=\frac{1}{2}$. To compute $P(4)$, first note that the last value must be greater than the previous one, so we choose from the four values two and put them in decreasing order. There are $\binom{4}{2}$ different alternatives. The two other non-selected values are put in the first two places, so there are 2! different configurations for them. Finally, the sequence (4,3,2,1) also verifies the condition, we have that $P(4)=(\binom{4}{2}2!+1)/4!=13/24$.


The following table presents the value $P(n)$ for larger values of $n$. For $n=\{6,7,8,9,10\}$ the exact values are computed and for larger values an estimate of $P(n)$ is presented by numerical simulations.
\begin{table}[h!]
	\begin{center}
		\begin{tabular}{|l|} \hline
			$P(6)=\frac{309}{720}\approx 0.429167$ \\
			$P(7)=\frac{2119}{5040}\approx 0.4204365$ \\
			$P(8)=\frac{16687}{40320}\approx 0.4138641$ \\
			$P(9)=\frac{148329}{362880}\approx 0.4087550$ \\
			$P(10)=\frac{1468457}{3628800}\approx 0.4046674$ \\
			$P(20)\approx 0.386271\pm 0.00097$ \\
			$P(50) \approx 0.375542\pm 0.00097$ \\
			$P(100) \approx 0.371565\pm 0.00097$ \\
			$P(200) \approx 0.36904\pm 0.00097$ \\
			$P(500) \approx 0.368494\pm 0.00096 $ \\
			$P(1000) \approx 0.367887 \pm 0.00096$ \\ \hline
		\end{tabular}
		\caption{$P(n)$ for different values of $n$. Exact values of $p(n)$ were calculated for $n=\{5,6,7,8,9,10\}$, and estimated values from simulations for $n=\{20,50,100,200,500,1000\}$. A 95\% confidence value is presented  for one million simulations.}
	\end{center}
\end{table}
Note that as $n$ grows $P(n)$ converge to the value $p_c$. For $n=1000$, we obtain a 95\% confidence interval of [0.366927,0.368847] for  $P(1000)$. Note that this interval contains $e^{-1} (\approx 0.3678794)$.

Next, we show that $p_c$ is in fact $e^{-1}$.
 A toy model for understanding self-organized criticality theory was introduced recently by Swart in the paper entitled ``A simple rank-based Markov chain with self-organized criticality''~\cite{swart}. This model belongs to the so called particles systems research area in probability theory. The model, as described in words of Swart, is the following:
``In each time step, a particle is added at a uniformly chosen position,
independent of the particles that are already present. If there are any particles to the left
of the newly arrived particle, then the left-most of these is removed.'' The author shows that
``all particles arriving to the left of $p^*_c = 1 - e^{-1}$ are a.s. eventually removed, while for large
enough time, particles arriving to the right of $p^*_c$ stay in the system forever.''

Interestingly, the model presented in~\cite{swart} can be reformulated in terms of our
model. In fact, this is the case if we replace the positions of the particles by
offers, and the leftmost by the highest (which is equivalent to the rightmost in
[51]). Therefore, the results of~\cite{swart} are valid for the sales model. Taking this
parallelism into account and with our notation, the main result (what was said
before in quotes) in~\cite{swart} is described in Theorem 1.


\begin{definition}\label{tau}
An avalanche is a sequence of events where the maximum remaining bid values, $X_{j}^{max}$, are greater than $e^{-1}$. It starts at $j+1$ if $X_{j}^{max} < e^{-1}$ and $X_{j+1}^{max}\geq e^{-1}$, and has a duration of $\tau$ if $X^{max}_{j+2}\geq e^{-1}$, $X^{max}_{j+3}\geq e^{-1}$,...,$X^{max}_{j+\tau}\geq e^{-1}$, and $X^{max}_{j+\tau+1}< e^{-1}$. Let $\tau_k$ be the duration of the $k$-th avalanche, and $\tau:=\underset{k \to \infty}\lim \tau_k$.
\end{definition}
\begin{theorem}[Swart 2017]
If $X_1, X_2, \dots$, is a sequence of iid random variables with Uniform[0,1] distribution then
\begin{align}\label{sss}
& \underset{k\to \infty}{\liminf} X_k^{max}=e^{-1}  \\
& \Pr{\tau< \infty}=1 \quad  \text{and} \quad \Ex{\tau}=\infty. \label{tau_swart}\\
& p_c=e^{-1} \label{c1} \\
& \underset{k\to \infty}{\lim} Y_k=:Y\sim \text{Uniform[$e^{-1}$,1]} \label{c2}\\
& \underset{k\to \infty}{\lim} Z_k=:Z\sim \text{Uniform[0,$e^{-1}$)}. \label{c3}
\end{align}
\end{theorem}



 On the following we study the random variable $X^{max}_k$ (which is the maximum remaining bid when $k$ bids have been offered) and its relationship with $p_c$. Let $H_k(x):=\Pr{X^{max}_k\leq x}$ be the distribution of $X^{max}_k$, and let $h_k(x)=\frac{\partial \Pr{X^{max}_k \leq x} } {\partial x}$ be the probability density function. And let call $U_{[a,b]}(x)=\Pr{U\leq x}$ the distribution of a Uniform[a,b] random variable $U$.
\begin{proposition}
If $X_1, X_2, \dots$, is a sequence of iid random variables with Uniform[0,1] distribution then
\begin{align}
p_s=\underset{k\to \infty}{\lim}\Ex{X^{max}_k}. \label{pps}
\end{align}
\end{proposition}

It is important to remark that $X^{max}_k$ does not converge to a Uniform distribution, $\underset{k\to \infty}{\lim} H_k(x)=:H(x) \neq U_{[e^{-1},1]}(x)$. And this can be understood by noticing that
 $\int^1_{e^{-1}} y \frac{1}{1-e^{-1}}dy\neq 1-e^{-1}=p_s$. The following Proposition presents the limit distribution of $X^{max}_k$.
\begin{proposition}
\begin{equation}\label{densh}
	\underset{k\to \infty}{\lim} h_k(y)=h(y)=\left\{
	\begin{array}{lll}
	\frac{1}{y} &  & \mbox{if}\ \  e^{-1} \leq y \leq 1 \\
      0 &  &  \mbox{otherwise}.
	\end{array}
	\right.
\end{equation}
\end{proposition}

\begin{figure}
	\begin{center}
\includegraphics[page=1,height=10cm,angle=0]{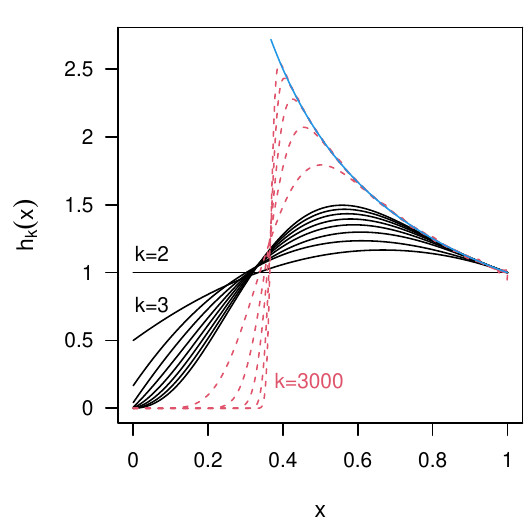}
	\caption{ Probability density function of $X^{max}_k$, $h_k(x)$ for different values of $k$. Black curves correspond to the exact pdf, for $k=2,3,\dots, 10$, and red curves to an estimation for $k$=30, 100, 300, 1000,and 3000. Each curve, from left to right, corresponds to increasing values of $k$. The horizontal line in panel corresponds to $k=2$, and the one most to the right to $k=3000$. The blue line corresponds to $h(x)$ given by eq.~\ref{densh}.} \label{fig:figh}
\end{center}
\end{figure}

For fixed $k$, the computation of $h_k(x)$ is difficult, but the firsts functions  $h_1(x), h_2(x), \dots, h_{10}(x)$  can be computed (see Supp. Mat. Sec. 1). Fig.~\ref{fig:figh} presents these first functions (black curves) together with estimations of $h_k(x)$ for large values of $k$ by simulations (red curves).  Note that for $k=3000$ the density almost has its support on $[p_c,1)$, and it is very similar to the asymptotic density ($h(x)$) given by eq.~\ref{densh} and plotted in blue color.

As we have explained, Theorem 1 ensures that when bids have a Uniform[0,1] distribution there is a sharp cut-off between accepted and non-accepted bids in the stationary regime. Non-accepted bids have support in the [0,$e^{-1}$) interval, accepted bids in [$e^{-1}$,1], the proportion of non-sales is equal to $p_c=e^{-1}$, and therefore the critical price $x_c=e^{-1}$. And, the asymptotic distribution of the maximum remaining bid ($\underset{k \to \infty}{lim} X^{max}_k$) has a pdf given by eq.~\ref{densh}.

 If we change the Uniform[0,1] bids distribution for an arbitrary distribution $F$ (with density $f$) the value $p_c$ remains unchanged and the same behavior is observed. This is because the sales depends only on the relative order between the bids.  But of course the critical price and the accepted and non-accepted bids distributions change. In this case, $x_c=F^{-1}(p_c)$ (eq.~\ref{xc}), and in the stationary regime the pdf of the non-accepted bids, $f^{na}$, becomes
\begin{equation}\label{densi1}
	f^{na}(z)=\left\{
	\begin{array}{lll}
		\frac{1}{F(x_c)} f(z)  &  &   \mbox{if}\ \ z < x_c \\
		0  &  & \mbox{if}\ \   z \geq x_c.
	\end{array}
	\right.
\end{equation}
I.e. $\Pr{Z\leq x}=\Pr{X\leq x| X<x_c}$. The accepted bids, $Y$, verify $\Pr{Y\leq x}=\Pr{X\leq x| X\geq x_c}$, or equivalently the pdf of $Y$, $f^{a}$, is
\begin{equation}\label{densi2}
	f^{a}(y)=\left\{
	\begin{array}{lll}
		0 &  &   \mbox{if}\ \ y < x_c \\
		\frac{1}{1- F(x_c)} f(y)  &  & \mbox{if}\ \   y \geq x_c.
	\end{array}
	\right.
\end{equation}
And the asymptotic distribution of $X^{max}_k$ verifies
\begin{equation}\label{densihf}
	 H(x)=\left\{
	\begin{array}{lll}
		0 &  &   \mbox{if}\ \ x < x_c \\
		\frac{1}{1- e^{-1}}\int^{x}_{x_c} \frac{f(y)}{y}dy  &  & \mbox{if}\ \   x \geq  x_c.
	\end{array}
	\right.
\end{equation}

\subsection{Purchase Price and Total Income}
The expected Total Income when $n$ bidders made their offers is
\begin{equation}\label{exact}
\Ex{TI(n)}=\Ex{\underset{Y\in \mathbb{Y}_n}{\sum} Y}.
\end{equation}
 In this section we are interested in the asymptotic value, the limit of the number of bidders $n \to \infty$, of the mean Income per bid, $\underset{n\to \infty}{\lim}\frac{1}{n} \Ex{TI(n)}$.

\begin{proposition}	
  \begin{equation}\label{asym}
  \underset{n\to \infty}{\lim}\frac{1}{n} \Ex{TI(n)}=\int^{\infty}_{x_c}xf(x)dx=p_s\Ex{X|X>x_c}.
\end{equation}
\end{proposition}

The histogram in Fig.1(A) shows in red the price probability density function (pdf) of the accepted bids, and the pdf of the non-accepted bids is depicted in grey. This simulation corresponds to the case where there are $k=1000$ bids. Note that, even for as few as 1,000 bidders, both distributions are almost the conditional distributions of eqs.~\ref{densi1} and~\ref{densi2}. There are only a few accepted bids below the critical value that are accepted at the beginning of the sales process (during the transient period). These few sales at the beginning, let us say there are $W$ sales below $p_c$, do not have any impact (become negligible) in the asymptotic ($k\to \infty$) distribution of sales/non-sales. This is true for any stationary process. What happens in the transient period is irrelevant for the asymptotic behavior, basically because $\underset{k\to \infty}{\lim}W/k=0$.

Therefore, the expected total income earned by the company when $n>>1$ interested buyers have made their bids can be calculated by
\begin{equation}\label{eti}
	\Ex{TI(n)} \approx n\int^{\infty}_{x_c}xf(x)dx.
\end{equation}
Unlike equation~\ref{asym}, equation~\ref{eti} is written for fixed $n$; although $n$ must be large, it is interesting to note the extent to which the description is good for small $n$.
Figure 1 (B) shows the $TI$ behavior as a function of the number of bidders for the sale process with a Log-Normal price valuation distribution. The empirical interval $TI \pm 3 \sqrt{Var(TI)}$ from 200 simulations is shown in red. More than 99\% of the simulations fall within this interval~\footnote{For large $n$, $TI$ converge by the Central Limit Theorem to a Normal distribution.}. The black line corresponds to the (theoretical) mean $TI$ value given by eq.~\ref{eti}.  A close-up of Fig. 1 for smaller values of $n$ is shown in the inset of Fig. 1 (B). This graph highlights the non-linear behavior at the beginning of the sales process, particularly when the number of bidders is less than 50. In this case, the number of non-sales is greater than $p_c$ as shown in Table 1. Also, since the set of frozen bids is very small, some bids with prices below the critical value are accepted. Therefore, less sales are concluded and less $TI$ is obtained. For $n$ larger than 50, eq.~\ref{eti} yields a good approximation. If we change the Log-Normal price valuation distribution by an exponential distribution with a rate of $\lambda$, then a simple expression can be obtained for the mean total income, $\Ex{TI(n)}\approx n(1-p_c)(1-\ln(1-p_c)).$ It is straightforward to compute the mean $TI$ for other price distributions such as Log-Normal, or heavy tail Power Law with $\alpha>2$. The case of $\alpha \leq 2$ is discussed in the Discussion section.

We have shown that if $n$ is large, the mean $TI$ is well described by eq.\ref{eti}. But, what about the variance of $TI$? How does it grow with $n$? Applying the same argument used for the expectation, it is straightforward to obtain that the variance verifies
\begin{equation*}
	\underset{n\to \infty}{\lim}\frac{1}{n}\Var{TI(n)}= a_f,
\end{equation*}
where $a_f=\int_{x_c}^{\infty}x^2f(x)dx-\left( \int_{x_c}^{\infty}xf(x)dx\right)^2$, and depends on the price distribution ($f$). This scaling behavior is good because for large $n$ values, it scales linearly with $n$, which is not much variability.  As what happens for the mean of $TI$, for small $n$ there will be some differences when we compare the empirical $TI$ variance with the theoretical variance, given by $\Var{TI(n)}=a_fn$ (only valid for large $n$). This difference between the variances is present for $N<50$, but the difference becomes negligible for larger $n$ values.

\subsection{Self-organized criticality and avalanches of sales}
The model presented here and its asymptotic behavior resemble the Bak-Sneppen model \cite{bak93,bak95,daniel}, which is well-known in the physics community. This model was introduced for modeling the evolution of species, and is one of the most elegant self-organized criticality (SOC) models. SOC models~\cite{soc,soc2,soc3,soc4} are a special type of dynamical model that have a (second-order phase transition) critical point as an attractor. These models display the behavior of systems posed in the critical point of phase transitions, but without the need for external tuning over a control parameter. That is to say, the system \textit{self-organized} around the critical point, which is why these models are called SOC. In general, SOC models are parsimonious models characterized by simple rules between agents/particles that give rise to emergent complex macroscopic behavior. These models display spatial and/or temporal scale-invariance properties observed in many natural and social phenomena. An analytical treatment of SOC models is difficult precisely because of the scale invariance properties. They are easily simulated on a computer, but only few exact results are known.

SOC theory is considered to be one of the mechanisms by which complexity arises in nature and social sciences, which is why the study of SOC theory and the development of new SOC models went beyond the domain of physics. Numerous mathematicians, economists, sociologists and biologists have been attracted by SOC. For instance, in economy, SOC has been used to model financial markets~\cite{finantial1,economy6}, the process of production and inventory~\cite{economy1,economy2}, innovation~\cite{economy3}, urban development~\cite{economy4}, and the effects of regulation on a market~\cite{economy5}, among other things. Here we present a SOC model for selling products in real time by an auction mechanism. Bidders compete and the system self-organizes in a situation where the (critical) sales price emerges. This critical price is a lower bound for buying the product; any bidder that offers a purchase value greater than this value will buy it (with probability one).

\begin{figure}
	\begin{center}
		\includegraphics[page=5,height=8cm,angle=0]{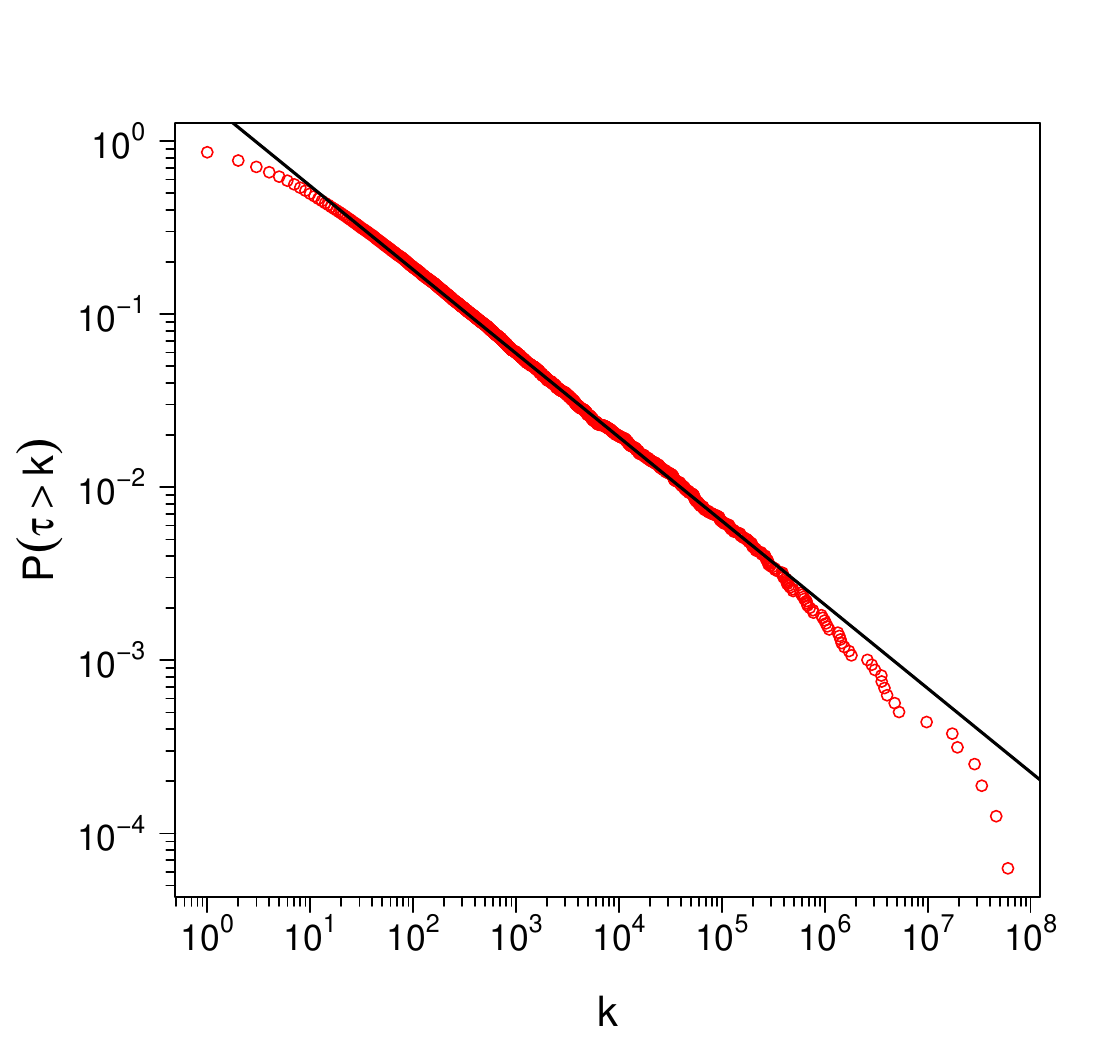}
	\caption{Avalanche distribution. Here we show the probability that the duration of a sales avalanche (of prices larger than $x_c$) is greater than an arbitrary value $k$ as a function of $k$, $P(\tau>k)$. The red line corresponds to a robust fit of the tail distribution in the range of $1000<k<10000000$. The estimated slope is $-0.49 \pm 0.02$.}  \label{avala}
\end{center}
\end{figure}
Moreover, SOC models are characterized by the presence of scale-free avalanches, which is a manifestation of the scale-invariance behavior of systems at the critical point. How about the current model? Indeed, the current model does contain scale-free avalanches (of sales). An avalanche, as defined in Def. 5, is a sequence of events where the accepted bid values are greater than $x_c$. It starts at $k+1$ if $Y_k < x_c$ and $Y_{k+1}>x_c$, and has a duration of $\tau$ if $Y_{k+2}>x_c$, $Y_{k+3}>x_c$,...,$Y_{k+\tau}>x_c$, and $Y_{k+\tau+1}\leq x_c$. If one studies the number of consecutive sales, $\tau$, above $x_c$ (or the number of sales between two successive purchases below price $x_c$), a power law behavior is observed (see Fig.~\ref{avala}). This behavior is consistent with Theorem 1 which states that $\Ex{\tau}=\infty$. Numerical simulations shows that
\begin{equation}
	P(\tau=s)\sim s^{-\alpha} \quad or \quad P(\tau>s)\sim s^{-\alpha+1},
\end{equation}
with $\alpha\approx 1.5$ (the same result is obtained by Swart also by simulations).

\subsection{Comparison with other sales mechanism}
In this section we compare the sales method proposed with other methods. It is important to mention that this comparison should not be considered conclusive, but rather a first approximation to the problem. The difficulty is that any sales mechanism determines the behavior of the bidders; in particular, it determines the distribution of prices $F(x)$, and the intention of buying (i.e., how many bidders, $n$, will make an offer and how much they will bid). These factors are more related to psychological issues than economic ones (considering the classic hypothesis that people maximize their utility). For example, it is well known that participative pricing mechanisms drive more sales (greater $n$) and this is very difficult to model in a realistic way. How much participation (which is an essential aspect for selling zero-marginal cost products) there will be in a particular participative sales mechanism can only be calculated by conducting an experiment or by an actual sale. Therefore, in this section, we will assume some hypotheses about bidders' behavior that are probably unrealistic, but that we believe are the least arbitrary. Having said that, below we attempt a comparison of our method with other sales mechanisms.

\subsubsection{Comparison with ``Pay what you want''.}

One might ask, why not sell to everyone who makes an offer (instead of defining an auction mechanism), as the band RadioHead did 10 years ago?  The band sold their album \textit{In Rainbows} using a ``pay what you want'' method that socializes the product. The music publishing company did not reveal details of this sale but admitted that more people downloaded the album for free than paid for it. A market research company estimated that only 38\% paid something~\cite{radiohead}, while 62 percent downloaded the album paying zero. In terms of bids, in a PWYW method all offered prices are accepted. Under this hypothesis, if we define the sequence of ``sure bids'', $X^{s}_1$, $X^{s}_2,  \dots, X^s_n$ these bids probably will be very different ($f^{s}(x)\neq f(x)$). It is reasonable to suppose that $X^{s}$ will be smaller than the value $X$ of the auction model presented here because there is no risk of not obtaining the product. Therefore, the total income will most likely be smaller with the PWYW (or donation) option. The conceptual argument that justifies this is that in the auction mechanism, the bidders are pressured to bid a ``reasonable'' price, since there is a chance of not obtaining the product they want. By contrast, in the PWYW method, the bidders know that they will obtain the product no matter what they bid and thus there is no stimulus for offering an ``interesting'' price.

The Total income for the PWYW method verifies,
$$TI^{s}(n)=\overset{n}{\underset{i=1}{\sum}}X^{s}_{1},$$
and its expectation $\Ex{TI^{s}(n)}=n\Ex{X^{s}}$.


For the auction method presented here,
$$TI(n)\approx \overset{n}{\underset{k=np_c}{\sum}}X_{(k)},$$
where $X_{(k)}$ correspond to the k-order-statistics.


If $X^{s}$ and $X$ have the same probability law, then clearly $\Ex{TI^{s}(n)}>\Ex{TI(n)}$. But the main point is that it is impossible that $X^{s}$ and $X$ have the same law. Any competitive mechanism will produce greater purchase prices. Therefore, it follows that
$$X^{s}\underset{st}{<}X \quad  \Longleftrightarrow \quad F(x)<F^s(x) \ \forall x,$$
because the proposed method ensures that the bidders ``make an economical effort'' to obtain the product. Now, depending on the exact differences between $F(x)$ and $F^{s}(x)$, the expected Total Income can be favorable for one mechanism or the other. At this point, we can only make assumptions about $F(x)$ and $F^{s}(x)$ to determine which sales mechanism is better.

The following example will be clarifier, let us suppose that in the PWYW method a proportion $p_0$ of the interested buyers pay zero, and a proportion $1-p_0$ offer something different from zero with law $F$ (and density $f$) and expected value $\mu$. And let us suppose that in the Auction model all bidders offer a price with the same law $F$, which is a conservative assumption~\footnote{Probably the distribution of prices for the auction is shift to the right in comparison with PYWY even for the ones that only offer a positive value.}. Then, for large $n$,
$$\Ex{TI^{s}(n)}=n\Ex{X}=(1-p_0)n\mu,$$
$$\Ex{TI(n)}\approx (1-p_c)n\mathbb{E}_{X\vert X>x_c}(X)=n\int^{\infty}_{x_c}xf(x)dx.$$
If we now suppose that $F$ is an exponential distribution (where exact calculations are straightforward) with rate $\lambda=\frac{1}{\mu}$ we obtain,
$\int^{\infty}_{x_c}xf(x)dx=  \mu (1+\frac{x_c}{\mu})exp(-\frac{x_c}{\mu})).$
Now, since $x_c=F^{-1}(1-p_c)$   ($x_c=-\mu \ln(p_c)$), the integral verifies
$\int^{\infty}_{x_c}xf(x)dx=\mu(1-\ln(p_c))p_c.$

Finally, the auction model will present greater Total Income if $p_0$ is smaller than $(1-\ln(p_c))p_c$. And using that $p_c=e^{-1}$, the Total Income of the Auction model will be greater than PWYW if
$$p_0>1-2e^{-1}\approx 0.2642.$$

As far as we know, the only available information about a real sale with the PWYW method is the case of Radiohead, where $p_0=0.62$ was reported, as mentioned above. Under this value of $p_0$, and if the exponential assumption distribution is true, then
$$\underset{n\to \infty}{\lim}\frac{\Ex{TI(n)}}{\Ex{TI^{s}(n)}}\approx 3.4.$$
I.e., the expected total income for the Auction method is 340\% greater than the one for PWYW.

The results above are valid under the hypothesis that bidders are not rational. If bidders make bids that maximize their expected payoff (see eq. 4 in Supp. Mat. Sec 4), then the Auction model is clearly better than the PWYW method. In the PWYW method, the expected payoff takes its maximum value when all participants bid zero ($g^{\star}_{Nash}(v)=0$). Therefore, if $n$ bidders bid, then the expected Total Income will be zero ($\Ex{TI^s(n)}=0$), which is smaller than the total income obtained in the auction model ($\Ex{TI(n)}\approx n (1-p_c)v_c$). In sum, it is likely that the auction model generates greater revenues than the PYWY method under both rational and non rational behaviors. Indeed, we conducted~\cite{marco2} several field experiments where we compared both mechanisms in true sales and observed that in all cases the auction mechanism presented here presents $\frac{TI}{TI^{s}}>3$. (i.e., the total income of the auction method was more than 300\% higher than the PWYW method).

\subsubsection{Comparison with Name your own price.}

As we mentioned, in the Name-Your-Own-Price (NYOP) sales technique the customer offers a price and if this price is higher than a threshold price ($x_{thld}$), fixed by the seller, then he/she gets the product at price $x$. The value $x_{thld}$ is unknown by the customer, and if $X$ is greater than $P$, then he/she buys the product and the seller gets an ``extra'' $x-P$ revenue.

In this case, it is very simple to calculate the expected Total Income when there are $n$ bidders,
\begin{equation}
\Ex{TI^{thld}(n)}=n \int^{\infty}_{x_{thld}} xf^{thld}(x)dx,
\end{equation}
where $f^{thld}(x)$ is the price pdf of the NYOP method. If we compare with the asymptotics of the method proposed in this paper, we obtain that if the sellers of both methods have the same price distribution, $f(x)=f^{thld}(x)$, which seems to be a very strong hypothesis, then
$$\underset{n\to \infty} {lim} \frac{1}{n}\Ex{TI^{thld}(n)} \leq \underset{n\to \infty} {lim} \frac{1}{n}\Ex{TI(n)},$$
if $x_{thld}>x_c$. For $x_{thld}=x_c$ we obtain the same asymptotic Total Income, and if $x_{thld}<x_c$ then the relation reverses. For small $n$, where the transient period is not negligible, the total income is $\Ex{TI^{thld}(n)}>\Ex{TI(n)}$ when $f(x)=f^{thld}(x)$. Nevertheless, as we mentioned, when selling a zero-marginal cost product the intention is to sell a large ($n>>1$) number of products. And therefore, the results for small $n$ are not very relevant.

Finally, note that in the NYOP method, the maximum Total Income is obtained for $x_{thld}=0$, from which we obtain the PWYW method. This result must be taken with caution because we are considering that $f^{thld}(x)$ does not change $x_{thld}$ (no information flow, as in a blind auction), and neither does the number of interested buyers.

Moreover, we think that the hypothesis $f(x)=f^{thld}(x)$, and the hypothesis that the number of buyers (or the rate of appearances) is equal, is not realistic. We believe there is a positive psychological effect that favors our proposed method. People like the idea that the whole community of interested buyers is in charge of fixing the critical price, as opposed to the idea that the seller unilaterally fixed it.

\subsubsection{Comparison with selling at a fixed price.}

In the auction model presented here, at the asymptotic limit, a bidder that offers a price $x$ greater than $x_c$ will get the product at the price offered ($x$). As mentioned, the value $x_c$ is an emergent of the selling rule (the seller does not intervene). In a traditional sales mechanism, the seller sells the product at a fixed price (all bidders that get the product pay the same) and this price is determined by the seller by maximizing the revenue. Let us define $x_{fix}$ as the optimal fixed price. All bid values that are at least $x_{fix}$ will be satisfied at price $x_{fix}$. Therefore, the mean Total Income when $n$ bidders have made their offers, $$\Ex{TI^{fix}(n)}=x_{fix}\Pr{X>x_{fix}}n.$$

\textbf{3.4.3.1 $F$ is known only by the seller.}

We will work under the same bidders' hypothesis: each bidder makes his/her own unique bid, $X$, without knowing either the previous bids or the bidders (i.e.,  $X_1$, $X_2$, $X_3$, $\dots, X_n$, is a sequence of i.i.d. random variables with probability density function $f(x)$ and cumulative probability $F(x)$. Bidders do not know the distribution $F(x)$, but, now we add the hypothesis that the seller know it.

 In order to maximize the total income, the optimal fixed price (also called Myerson reserve price) must verify
\begin{equation}\label{fix}
	x_{fix}=\underset{x\in \R^+}{argmax} \ \ x \cdot \Pr{X>x}=\underset{x\in \R^+}{argmax} \ \ x \left(1-F(x)\right).
\end{equation}
Eq.~\ref{fix} shows that $x_{fix}$ depends on the bid probability law ($F(x)$). For example, if the bids $X's$ have an exponential law with rate $\lambda$, then optimal fixed price is $x_{fix}=1/\lambda$ and the mean total income, $\Ex{TI^{fix}(n)}=\frac{1}{\lambda}e^{-1}n$. For large $n$ both quantities $x_c$ and $\Ex{TI^{fix}(n)}$ can  be calculated for our model. In this case, $x_c=-\frac{1}{\lambda}ln(1-p_c)$, and by eq.~\ref{asym}
\begin{equation*}
	\Ex{TI(n)}\approx n \int^{\infty}_{x_c} \lambda x e^{-\lambda x}dx=n\frac{\lambda x_c +1}{\lambda}e^{-\lambda x_c}.
\end{equation*}
Replacing $x_c$ in the previous equation, $\Ex{TI(n)}\approx \frac{(1-ln(1-p_c))(1-p_c)}{\lambda}n$. And since $p_c=e^{-1}$, we obtain $\Ex{TI}\approx \frac{1}{\lambda}0.92206n$. Note that the total income of our model is 2.5 times greater than one from the fixed price strategy. This comparison is presented in the following table, which also shows the results for a Uniform distribution.\begin{table}[h!]
	
	\begin{center}
			\begin{adjustbox}{max width=0.9\textwidth}	
		\begin{tabular}{|c|c|c|c|c|c|}
			\hline
			
			 & \multicolumn{2}{c|}{Fixed price} & %
			\multicolumn{2}{c|}{Our Model}& Comparison \\
			\cline{2-6}
			\cline{2-6}
			\rule{0pt}{0.5cm}
			$F(x)$& $x_{fix}$ & $\Ex{TI^{fix}(n)}$ & $x_c$  & $\Ex{TI}(n)$ & $\underset{n\to \infty}{\lim}\Ex{TI(n)}/\Ex{TI^{fix}(n)}$  \\
			\hline
			 Uniform(0,1) & 0.5 & $0.25 n$ & 0.36787 & $0.4323 n$ &  1.7293 \\
			\hline
			 Exponential($\lambda$)& $1/\lambda$  & $0.36787n/\lambda$ & $0.45868/\lambda$& $ 0.92206n/\lambda$ & 2.50648 \\
			\hline
		\end{tabular}
\end{adjustbox}
		\caption{Comparison of the mean Total Income and the price values ($x_f$ and $x_c$) for the optimal fixed price model and our model.}
	\end{center}
\end{table}

	It is difficult to determine the exact condition for $\Ex{TI(n)}/\Ex{TI^{fix}(n)}$ as being greater than one. The following modest result shows a characterization of $f(x)$ from which we can assert that this quotient is greater than one in the asymptotic limit.

\begin{proposition}
	If the bids price distribution is continuous with decreasing and differentiable density, $f(x)$, and $f(x)<\frac{1-p_c}{x}$ then $Q:=\underset{n\to \infty}{\lim}\Ex{TI(n)}/\Ex{TI^{fix}(n)}>1$.	
\end{proposition}	

There are also other $f(x)$, in particular non-decreasing ones, that present a total income $\underset{n\to \infty}{\lim}\Ex{TI(n)}/\Ex{TI^{fix}(n)}>1$. Next, we analyze three other families of distributions that are very flexible for describing data: Gamma($\alpha, \lambda$), Beta($\alpha, \beta$) and Normal($\mu,\sigma$) distributions.

\begin{figure}
	\begin{center} 
		\includegraphics[height=8.5cm,angle=0]{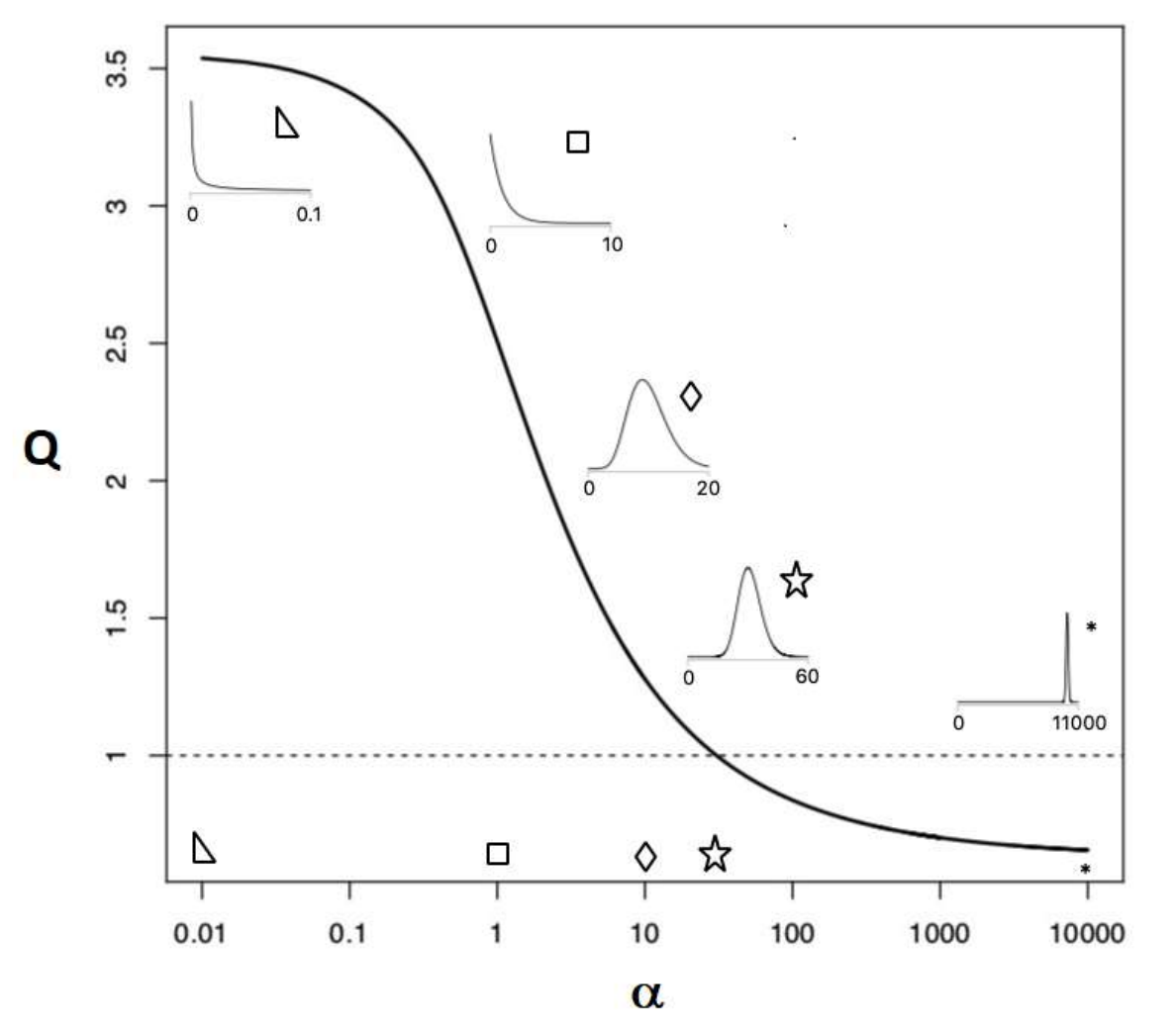}
\end{center}
\caption{$Q$ as a function of the shape parameter $\alpha$ of a Gamma distribution with $\lambda=1$. The probability density function, $f(x)$, is shown for some values of $\lambda$.}
\end{figure}
Figure 3, and supplementary figures 1 and 2 present the quotient $Q$ for each family. Fig. 3 shows the results for the Gamma($\alpha, \lambda$) distribution. It can be noted that for $\alpha$ smaller than approximately $31$ our model yields higher expected Total Income than the fixed price method. For very small values of $\alpha$ our model yields 3 times the TI of the fixed priced mechanism. For $\alpha$ greater than $31$ the behavior reverts. These results correspond to the scale parameter $\lambda=1$, but equal results are obtained for all values of $\lambda$ (data not shown) (i.e., the scale does not play a role in the quotient $Q$). In the inset of the figure, the density, $f(x)$, is shown for some values of $\alpha$ which are denoted by geometric symbols.

Supp. Mat. shows the behavior of $Q$ when the bid price distribution is Normal, and also when is Beta($\alpha$, $\beta$).

\textbf{3.4.3.2. $F$ is known by bidders and the seller.}

If $F$ is known by bidders and the seller, it is expected that bidders will use an optimal strategy and therefore offer $x_c$ as maximum value (see eq. 6 in Supp. Mat. Sec 4). In this case, the total income verify
 $$\Ex{TI^{fix}(n)}= x_{fix}P(X>x_{fix})n=\underset{x\in \R}{\arg\max} \ xP(X>x)n >  \Ex{TI(n)}\approx x_cP(X>x_c)n.$$
 For example, if $F(x)$ is Uniform[0,1] then $x_{fix}=1/2$ and the Total Income for $n$ ``bidders'' is $\Ex{TI^{fix}(n)}=1/2\cdot 1/2 \cdot n$. And for the method proposed here, $x_c=e^{-1}$ and the total income is $\Ex{TI(n)}\approx e^{-1}\cdot(1-e^{-1})\cdot n$. Note that the calculus consider that the number of interested buyers ($n$) is the same in both methods. But, it has been reported~\cite{Pay9} that participative pricing mechanism drive more buyers. So, an experiment must be done to really know if
$\Ex{TI^{fix}(n_1)}$ is greater o smaller than  $\Ex{TI(n_2)}$ when $n_2>n_1$.

\textbf{3.4.3.3. $F$ is unknown by bidders and the seller.}

When $F$ is unknown by bidders and the seller it is not possible to calculate the $x_{fix}$, but under additional information, such as the mean of the distribution, the worst-case criterion sometimes can be calculated (see for example~\cite{worst_case}). Here without assuming any additional information,
 we present a general result that favors the fixed price method.
\begin{definition}
The mean Total Income of worst case is defined as
$$\Ex{TI_{WC}(n)}:=\underset{x \in \R^{+}}{sup} \underset{F \in \mathbb{F}}{inf} \Ex{TI(n, F, x)},$$
where $\mathbb{F}$ is the family of distributions. 
\end{definition}

As before, we use the sub/superscript \textit{fix} to denote the results of the model with a fixed price ($TI_{WC}^{fix}$, $x_{fix}$).
\begin{proposition}
If F is unknown for bidders and the seller, then
\begin{align*}
& \Ex{TI_{WC}(n)} < \Ex{TI^{fix}_{WC}(n)}, \\
&  \underset{n\to \infty} {\lim} \frac{\Ex{TI_{WC}(n)}}{\Ex{TI^{fix}_{WC}(n)}}\leq 1.
\end{align*}
\end{proposition}

But as we mentioned before, it is important to remark that this comparison is not completely correct. We have assumed that $n$ is the same for both methods (here we are not assuming that $F$ is the same), and this seems to be false in practical implementations~\cite{Pay9}.

\subsection{Strategic behavior of bidders.}
Bidders appears when they become interested in the product and are able to make an offer. In the model, this is represented by independent random price bids ($X$'s) that appear at different random times. The results presented above are valid for all types of time appearance processes, including determinist ones (e.g., one bid per minute). Only the order of the bids (and not the appearance times) is relevant for the model. In the model it is assumed that all bidders are interested in the product, and the independent price valuations among bidders is justified by the fact that it is a Blind auction.

It is important to mention that the model is robust against the strategic behavior of bidders or collusion.  Bidders cannot collude by agreeing on bidding prices (generating dependence between some $X$'s). If they agree with a pricing strategy (e.g. the first bidder offers a low price and immediately a second bidder offers an even lower value) neither of them will obtain the product. As we have shown, the product is obtained only by bidders who have offered a price above the critical one. Solely at the start of the auction (transient behavior) can bidder collusion take place. Nevertheless, the impact of these potential few low price sales will be negligible. Moreover, this small risk can disappear if the seller waits some time, accumulating bidders, and the bidders are reshuffled (random order) before accepting the first auction with the model's rule.

\section{Discussion}
It is well established that participatory pricing increases consumers' intent to purchase, and can therefore be an opportunity for companies to surpass the competition. Here we have introduced a novel participative pricing mechanism for selling products in real time. This mechanism exhibits self-organized criticality~\cite{soc}. The model can be applied to sell any product that has infinity stock or products that can be produced at the same (or similar) rate as the demand (e.g., online ads, electronic posters, software, etc.). In the model for almost all bids, the decision is made quickly and the acceptance price is above a (critical) value that only depends on the valuation distribution. Approximately 63.2\% ($(1-e^{-1})100\%$) of the bidders will buy the product (bids accepted).  One of the advantages of the model presented here is that the average total income can be estimated with high accuracy. And as we have shown, under both rational and not rational hypotheses, it is likely that this method yields a larger total income compared with the PWYW method, and probably also with NYOP. We also compared the method with a fixed price strategy. In this case, the results depend on the shape of the bid distribution.

It is interesting to note that the selling context analyzed here, where the seller has infinite products to sell, is very different from the traditional one where the seller only has one product to sell. In this last case, there are many mathematical techniques that study which auction rules lead to the highest average prices. However, in our case, this is still missing. We believe that the model presented here may motivate quantitative researchers to further explore this topic and develop new models for continuous sales of infinite products. Moreover, behavioral economic experiments and empirical data may help us answer:  which auction rules lead to the ``highest'' total income?, and more precisely, how is $F(x)$ modified by the auction rule? This last question is especially important for identifying a true optimal auction mechanism.

Finally, we believe that our model could also be used by companies to establish the initial price of a novel product with limited or unlimited stock.

In the following subsections, we discuss some considerations for implementing the model and some relevant questions: What is the empirical price distribution? Is it convenient to set a base price?

\subsection{Considerations for the implementation of the model}
One important consideration for implementing the model is that each interested buyer must apply to the auction process through a \textit{unique} bid, as in closed envelope tenders. Therefore, in practice, it is necessary to verify bidder identities, which can easily be done when the sales product is a smartphone app.

\subsubsection{Modifications of the model.}
The model may be modified to have a different critical value and similar dynamics. For example, a bid would be accepted if it is the highest bid compared with those presented before and after it \textit{two consecutive}. This last model yields a critical value $p_c\approx 0.5$.

Another modification of the original model may be to use the same rule and add the possibility that bidders at any time may request to withdraw from the auction. If the seller selectively decides which request is accepted and at what moment the withdrawal occurs, then the model can have exactly the same behavior. Once there are a large number of bids, the seller may withdraw the frozen bids. However, the seller must make sure that the total rate of accepted withdrawals be much smaller than the rate of appearance of new bids, thus making sure the withdrawals are substituted by the new similar bids that appear. Under these conditions, the frozen distribution and the active distribution will be as stable as the ones described in eqs.~\ref{densi1} and~\ref{densi2}. The obvious question now is, can the withdrawn bidders buy the product on a second occasion when they are really interested in it? In this case, based on the history of accepted bids, the seller needs to offer the product to each withdrawer at a different random price selected from the history of accepted bids. But, offering this second chance to buy the product has some consequences. On the one hand, bidders will be happy because they get a second opportunity to obtain the product, and on the other hand, it is likely that the first price bid, $X^{first}$, will be stochastically smaller than the bid $X$ when there is no second opportunity. Therefore, the critical price in this case will be smaller than in the original case where there is only one chance to buy. Finally, the mean income per accepted bid will be smaller, but the proportion of sales will likely be greater. We do not have a way to check which of both total incomes will be greater. A controlled experiment must be conducted to measure: (A) the proportion of bidders that are interested in a second offer, (B) the proportion of bidders that truly buy at the random price offered by the seller, (C) the price distribution with both methods, and (D) the number of bidders (or rate of appearance) of both methods. However, these results may only be valid for the product offered. In other words, changing the product may lead to different results. But, under the rationality hypothesis (utility maximization) bidders must offer a very small first bid ($0<\epsilon <<1$, bids must be positive) such that the second price offered by the seller is very low. Therefore, it does not seem to be a good option to offer a second chance of buying. Second chances could be offered occasionally and not be included in the selling rule.

Finally, another modification to the model can be to allow bidders to withdraw after some time, more precisely after $k_{lag}$ new bids appear. But, taking into account that there is no second chance of buy, and that the bidders can withdraw without an expressed authorization of the seller, in this case, we can obtain exactly the same behavior of the original model or a similar behavior. This model exceeds the scope of this paper and will be presented elsewhere.

Shortly, depending on the parameters $k_{lag}$ and the probability that a bidder withdraw, the model can be equal to the original model with the same sharps distributions for accepted and frozen bids and the same critical price. Or can be similar with smoothers distributions of accepted and non-accepted bids and without a clear cutoff and a critical price. Namely, the abrupt transition between buying and being frozen will be lost,
and therefore we will not have a well-defined critical price. In this case we do not obtain a self-organised criticality model.  The larger $k_{lag}$ the more similar the behavior of this model to the original.

\subsubsection{The price distribution.}
Clearly, the bidder's behavior is influenced by his/her income, the opportunities presented by the economic environment, the valuation of the product, and the auction's rules, among other things~\cite{sub0}. That is why bidders present different bids. If we randomly choose one bidder, she/he will make an economic offer that is described by a random variable $X$ with distribution $F(x)$. The company profit depends on this last distribution $F(x)$, which is a function of the valuation distribution $F_V(v)$. We cannot advocate for any one valuation distribution over another, but suggest that novel products or services would have an exponential or power law distribution, while products that are well-known on the market would have a Normal or Log-Normal distribution. It is interesting to note that if the price distribution has a power law tail with $\alpha < 2$, the expected total income is infinity, which means in practice, it could be arbitrarily large.

Is there a universal valuation distribution for novel products? In other words, is the law the same for all products, respecting a given scale factor? One way to determine this is to use the model on real sales. Behavioral experiments could help answer this question if the true motivation for obtaining the product is controlled, which is difficult to do. Finally, one should consider the wealth distribution in a given population, which would reflect that population's economic capacity to buy the product. This may ``mix'' the ``true'' price distribution of a homogeneous population. Perhaps all of these factors together result in a heavy tail distribution. It would be interesting to conduct studies in different countries (i.e., using different GPS coefficients).

\subsubsection{Targeting prices.}
With additional information about the bidders, one can categorize them according to country, sex, age, and any other relevant sociodemographic variable. Bidders could compete with other bidders from the same economic segment, which would yield more equity opportunities for acquiring the product. The $(1-p_c)100$\% most interested targeted buyers (based on the bids) will obtain the product.

\subsubsection{Base price or not?}
The same auction procedure may be applied with a base price. In this case, the bids received ($X$) will be larger than the base price. Companies that sell non-zero products may be tempted to use a base price. Is this a good strategy? Will the profit be larger? This is not an easy question to answer. Once again, this most likely depends on the novelty of the product. However, setting a base price may have a priming effect~\cite{priming3}, a phenomenon well-known to the cognitive neuroscience and behavioral economics communities. In priming~\cite{priming1,priming2}, exposure to a stimulus influences a response to a subsequent stimulus without conscious guidance or intention. How much awareness the bidder has regarding the priming effect on his/her bid value is a matter of debate.

\section*{Appendix: Proofs of Statements.}

\begin{proof}{\bf \emph{Proposition 1.}}
By Definition 4,
\begin{align*}
p_s &=\underset{k\to \infty}{\lim}P(k+1)=\underset{k\to \infty}{\lim}\Pr{X_{k+1}< X_k^{max}}= \underset{k\to \infty}{\lim}\int^{\infty}_0 P(X_{k+1} <y|X^{max}_k=y)h_k(y)dy\\
&= \underset{k\to \infty}{\lim}\int^{\infty}_{0} yh_k(y)dy.
\end{align*}
The pdf $h_k(x)$ converges to a pdf $h(x)$ which has support in $[e^{-1},1)$ (by Theorem 1), and therefore we obtain eq. 21.
\qed
\end{proof}

\begin{proof}{\bf \emph{Proposition 2.}}
Let $A^{\uparrow}_k$ be the set of non-accepted bids with values greater or equal $e^{-1}$ when $k$ bids have been offered. The set $A^{\uparrow}_1$ has the element $X_1$ with probability $1-e^{-1}$ or it is empty set. Let $W_k:=max\{A^{\uparrow}_k\}$ be the maximum value of the set, and in order to have a more parsimonious notation let define $\max\{ \emptyset \}=0.$ The evolution of $A^{\uparrow}_k$ is the following,
\begin{equation}\label{A}
	A^{\uparrow}_{k+1}=\left\{
	\begin{array}{lll}
      	A^{\uparrow}_k &  & \mbox{if}\ \   (X_{k+1}<e^{-1}) \cap (W_k <e^{-1}) \\
		A^{\uparrow}_k\cup \{X_{k+1}\} &  & \mbox{if}\ \  (e^{-1} \leq X_{k+1})\cap (W_k \leq X_{k+1}) \\
      A^{\uparrow}_k\cup \{X_{k+1}\}\setminus \{W_k\} &  & \mbox{if}\ \   e^{-1} \leq X_{k+1} < W_k\\
        A^{\uparrow}_k\setminus \{W_k\} &  & \mbox{if}\ \   (X_{k+1} \leq e^{-1}) \cap (e^{-1}<W_k).
	\end{array}
	\right.
\end{equation}

We are going to study $X^{max}_{k}$ in the stationary regime ($k>>1$). In this regime, we know that
there exist an accumulation of non-accepted or frozen bids that are Uniformly distributed on [0,$e^{-1}$).
Let $k_1>k$ be the first value that verify $X^{max}_{k_1-1}<e^{-1}$ ($|A^{\uparrow}_{k_1-1}|=0$), and $X^{max}_{k_1}=x \geq e^{-1}$ ($|A^{\uparrow}_{k_1}|=1$).   Note that $k_1$ now is a random variable. And let call $\tilde{k}_1$ the first following value that verify $X^{max}_{\tilde{k}_1}<e^{-1}$, and $\tau_1:=\tilde{k}_1-k_1$. The random variables $k_2, \tilde{k}_2, k_3, \tilde{k}_3, \dots$, are defined in the same way.

First note that for studying the asymptotic of  $X^{max}_{k}$ the values that are smallers than $e^{-1}$ can be neglected because $\Ex{\tilde{k}_i-k_i}=\infty$ (see eq. 18 in main text) and $\Ex{k_{i+1}-\tilde{k}_i}=d<\infty$ (in fact, $d=1/(1-e^{-1})$).  Therefore,  instead of studying the complete sequence $X^{max}_{k_1},X^{max}_{k_1+1},\dots$, we will study the sequence $$X^{max}_{k_1},X^{max}_{k_1+1},\dots,X^{max}_{\tilde{k_1}-1},X^{max}_{k_2},X^{max}_{k_2+1},\dots, X^{max}_{\tilde{k}_2-1},X^{max}_{k_3}, \dots$$.

Now, note that the above sequence have repeated values, for example, a sequence that can be observed is $X^{max}_{k_1}=x, X^{max}_{k_1+1}=x+\delta$ (with $0<\delta<1-x$), and $X^{max}_{k_1+2}=x$. Therefore, in order to understand the asymptotic distribution of $X^{max}_k$ we need to understand the multiplicity of each value $X^{max}_{k}$, or number of times a value is repeated.


Based on eq.~\ref{A} and noticing that $W_k=X^{max}_k$ when $X^{max}_k\geq e^{-1}$, it is easy to check that
\begin{equation}
\Pr{x\notin A^{\uparrow}_{k+1}|X^{max}_k=x}=x \quad  \forall k\in \N \ \text{if} \ x\geq e^{-1}.
\end{equation}
And the complement
\begin{equation}\label{probaa}
\Pr{x\in A^{\uparrow}_{k+1}|X^{max}_k=x}=1-x \quad \forall k\in \N \ \text{if} \ x\geq e^{-1}.
\end{equation}
  Now, let define the set of ``time-events'' where $x$ is the maximum bid, $$S_x=\{k_1 \leq j<\tilde{k}_1: X^{max}_j=x\}.$$
The random variable $n_x:=|S_x|$ is the multiplicity of the value $x$, and it is easy to check that the multiplicity is zero if $x$ has not appeared or has a Geometric distribution with parameter $p=x$ if it has appeared. Then if $x$ appear, the multiplicity expectation is $\Ex{n_x}=\frac{1}{x}$. Finally,
\begin{align}
  H(x):&=\underset{k\to \infty}{\lim} \Pr{X^{max}_k\leq y} =
    \frac{1}{\int^{1}_{0} f^a(w)\Ex{n_w}dw}\int^{y}_{0} f^a(x)\Ex{n_x}dx\\
  &=(1-e^{-1})\int^{y}_{0} f^a(x)\Ex{n_x}dx\\
   &=\left\{
	\begin{array}{lll}
	0 &  & \mbox{if}\ \   y < e^{-1} \\
     \int^{y}_{e^{-1}}\frac{1}{x}dx &  &  \mbox{if}\ \  e^{-1} \leq y \leq 1\\
      1&  & \mbox{if}\ \ y>1.
	\end{array}
	\right.
=\left\{
	\begin{array}{lll}
	0 &  & \mbox{if}\ \   y < e^{-1} \\
      \ln(y)+1 &  &  \mbox{if}\ \  e^{-1} \leq y \leq 1\\
      1&  & \mbox{if}\ \ y>1.
	\end{array}
	\right.
\end{align}
Or it's density $h(y)=\frac{\partial H(y)}{\partial y}$ verify eq. 22.
\end{proof}

\begin{proof}{\bf \emph{Proposition 3.}}
\begin{align*}
  \underset{N\to \infty}{\lim}\frac{1}{n} \Ex{TI(n)} &= \Pr{\text{accept a new bid}}\Ex{Y}
  = p_s\int_{0}^{\infty}yf^{a}(y)dy=\int_{x_c}^{\infty}yf(y)dy\\
&=p_s\Ex{X|X>x_c}.
\end{align*}
\qed
\end{proof}

\begin{proof}{\bf \emph{Proposition 4.}}
	\begin{align}\label{xf}
		& \frac{\partial}{\partial x} \left(\ x\cdot\Pr{X>x} \right)=1-F(x)-xf(x)=0 \leftrightarrow \\
		\nonumber & F(x_{fix})=1-x_{fix}f(x_{fix})
	\end{align}
The value $x_{fix}$ is a maximum,  $\frac{\partial^2 }{\partial x^2}\left(\ x\cdot\Pr{X>x} \right)=\frac{\partial f(x)}{\partial x}<0$,
since by hypothesis $f(x)$ is decreasing. Now, if $x_{fix}>x_c$, which is equivalent to $F(x_{fix})>p_c$, or to
	\begin{equation}\label{xxf}
		f(x_{fix})<\frac{1-p_c}{x_{fix}},
	\end{equation}
we can assert that $\Ex{TI}>\Ex{TI^{fix}}$. Finally, we impose by hypothesis that eq.\ref{xxf} is valid for all values of $x$ to ensure that there is a solution that verifies both conditions (eq.~\ref{xf} and~\ref{xxf}). 
\qed
\end{proof}

\begin{proof}{\bf \emph{Proposition 5.}}
For the model presented here remember that $x_c=F^{-1}(p_c)$ and that for large $n$ all bids greater than $x_c$ are accepted. Therefore, $\Ex{TI(n)}\approx F^{-1}(p_c)(1-p_c)n$ (or for small $n$ $\Ex{TI(n)}<F^{-1}(p_c)(1-p_c)n$) which entails,
\begin{equation}
  \Ex{TI_{WC}(n)}<\underset{F \in  \mathbb{F}}{inf}   F^{-1}(p_c)(1-p_c)n\leq \underset{x \in \R^{+}}{sup} \underset{F in \mathbb{F}}{inf} x(1-F(x))=\Ex{TI_{WC}^{fix}(n)}. \\
\end{equation}
The last inequality holds because $\Ex{TI_{fix}^{WC}(n)}$ has a supremum over $x$. And in the limit $n \to \infty$ we have $\underset{n\to \infty}{\lim} \frac{1}{n} \Ex{TI_{WC}(n)}=\underset{F \in  \mathbb{F}}{inf}  F^{-1}(p_c)(1-p_c)$.
\qed
\end{proof}

\bibliographystyle{nonumber}

\begin{thebibliography}{11}
	\bibitem[Krishna 2009]{auction} Krishna, V. 2009.   Auction theory. \textit{Academic press}.
	
	\bibitem[Gallego and Van Ryzin 1994]{gallego} Gallego, G.,  Van Ryzin, G. 1994. Optimal dynamic pricing of inventories with stochastic demand over finite horizons. \textit{Management science}, 40(8), 999-1020.
	
	\bibitem[Myerson 1981]{sub1} Myerson, R. 1981. Optimal Auction Design.   \textit{Mathematics of Operations Research}, \textbf{6},
	58-73.
	
	\bibitem[Riley and Samuelson 1992]{sub2} Riley, J., Samuelson, W. 1982. Optimal Auctions. \textit{American Economic Review}, \textbf{71},
	381-392.
	
	\bibitem[Klemperer 2004]{sub3} Klemperer, P. 2004  Auctions: Theory and Practice. \textit{Princeton University Press}.
	
	\bibitem[Engelbrecht-Wiggans 1993]{sub4} Engelbrecht-Wiggans, R. 1993. Optimal Auction Revisited. \textit{Games and Economic Behavior},
	\textbf{5}, 227-239.
	
	
	\bibitem[Lahaie et al. 2007]{publicidad}  Lahaie, S., Pennock, D.M., Saberi, A., and Vohra, R. V. 2007. Sponsored search auctions.
	\textit{Algorithmic Game Theory},  699-716.
	
	
	\bibitem[Goldberg et al. 2006]{goldberg} Goldberg, A., Hartline, J., Karlin,A., Saks, M.,  Wright, A. 2006. Competitive auctions. \textit{Games and Economic Behavior}, 55(2), 242-269.
	
	\bibitem[Goldberg et al. 2001]{digital} Goldberg,A., Hartline,J., Wright, A. 2001. Competitive auctions and digital goods. \textit{Proc. 12th Annu. ACM-SIAM Symp. Discrete Algorithms (SODA)},  735-744.
	
	
	\bibitem[Just and Wansink 2011]{Flatrate1}  Just,D., Wansink, B. 2011. The flat-rate pricing paradox: conflicting effects of ``all-you-can-eat'' buffet pricing.  \textit{The Review of Economics and Statistics},  \textbf{93}(1), 193-200.
	
	\bibitem[Hamari et al. 2017]{Freemium1} Hamari,  Hanner,J. N.,   Koivisto,J. 2017. Service quality explains why people use freemium services but not if they go premium: An empirical study in free-to-play games.  \textit{International Journal of Information Management},  \textbf{37}(1), 1449-1459.
	\bibitem[Kanna 2017]{Freemium2}  Kannan, P. 2017. Digital marketing: A framework, review and research agenda.  \textit{International Journal of Research in Marketing},  \textbf{34}(1), 22-45.
	\bibitem[Shampanier et al. 2007]{Freemium3} Shampanier, K.,  Mazar, N., Ariely, D. 2007. Zero as a special price: The true value of free products.  \textit{Marketing science},  \textbf{26}(6), 742-757.
	\bibitem[Gu et al. 2018]{Freemium4} Gu, X., Kannan, P.,   Ma,  L. 2018. Selling the Premium in Freemium.  \textit{Journal of Marketing},  \textbf{82}(6), 10-27.
	
	\bibitem[Pauwels and Weiss 2008]{Freemium5}  Pauwels, K., Weiss, A. 2008. Moving from free to fee: How online firms market to change their business model successfully.  \textit{Journal of Marketing},  \textbf{72}(3), 14-31.
	
	\bibitem[Bormann 2017]{Freemium6}  Bormann, P. 2017. Freemium or paid? The Impact of firm experience and app data on the revenue streams of mobile gaming apps.
	
	\bibitem[Boss 2018]{Freemium7} Ross, N. 2018. Customer retention in freemium applications  \textit{Journal of Marketing Analytics} \textbf{6.4}, 127-137.
	
	\bibitem[Clarence et al. 2015]{Freemium8}  Clarence,L., Kumar, V.,  Gupta, S. 2015. Designing freemium: Balancing growth and monetization strategies Available at SSRN.
	
	
	\bibitem[Chandran and Morwitz 2005]{Pay9}  Chandran,S.,  Morwitz, V. 2005. Effects of Participative Pricing on Consumers' Cognitions and Actions: A Goal Theoretic Perspective,  \textit{Journal of Consumer Research}, \textbf{32} (2), 249-59.
	
	\bibitem[Roy et al. 2016]{Pay8} Roy, R.,Rabbanee, F.,  Sharma, P. 2016. Exploring the interactions among external reference price, social visibility and purchase motivation in pay-what-you-want pricing.  \textit{European Journal of Marketing}, \textbf{50}(5/6), 816-837.
	
	\bibitem[Roy 2015]{Pay7}  Roy, R. 2015. An insight into pay-what-you-want pricing.  \textit{Marketing Intelligence \& Planning}, \textbf{33}(5), 733-748.
	
	\bibitem[Kim et al. 2014]{Pay3}  Kim, J.,  Kaufmann, K.,  Stegemann, M. 2014. The impact of buyer seller relationships and reference prices on the effectiveness of the pay what you want pricing mechanism.  \textit{Marketing Letters}, \textbf{25}(4), 409-423.
	
	\bibitem[Schmidt et al. 2014]{Pay1}   Schmidt,K.,  Spann,M.,  Zeithammer, R. 2014. Pay what you want as a marketing strategy in monopolistic and competitive markets.  \textit{Management Science}, \textbf{61}(6), 1217-1236.
	
	\bibitem[Kim et al. 2009]{Pay2}  Kim,J.,  Natter, M., Spann,  M. 2009. Pay what you want: A new participative pricing mechanism.  \textit{Journal of Marketing}, \textbf{73}(1), 44-58.
	
	\bibitem[Schons et al. 2014]{Pay4}  Schons,L., Rese,M. ,  Wieseke,J.,   Rasmussen,W.,  Weber,D., Strotmann, W. 2014. There is nothing permanent except change-analyzing individual price dynamics in ``pay-what-you-want'' situations.  \textit{Marketing Letters}, \textbf{25}(1), 25-36.
	
	
	\bibitem[Jang and Chu 2012]{Pay5}  Jang,H.,  Chu,W. 2012. Are consumers acting fairly toward companies? An examination of pay-what-you-want pricing.  \textit{Journal of Macromarketing}, \textbf{32}(4), 348-360.
	
	\bibitem[Chao et al. 2015]{Pay6}  Chao,Y.,  Fernandez,J.,    Nahata,B. 2015.  Pay-what-you-want pricing: Can it be profitable?.  \textit{Journal of Behavioral and Experimental Economics}, \textbf{57}, 176-185.
	

\bibitem[Bak and Snepen 1993]{bak93} Bak, P.,  Sneppen, K. 1993. Punctuated equilibrium and criticality in a simple model of evolution. \textit{Phys. Rev. Lett.}, \textbf{71}, 4083.
	
	\bibitem[Sneppen et al. 1995]{bak95} Sneppen, K.,  Bak, P.,  Flyvbjerg, H., Jensen, M. 1995. Evolution as a self-organized critical
	phenomenon. \textit{Proc. Natl. Acad. Sci. USA}, \textbf{92},  5209.
	
	\bibitem[Fraiman 2018]{daniel}  Fraiman, D. 2018. Bak-Sneppen model: Local equilibrium and critical value. \textit{Phys. Rev. E}, \textbf{97}, 042123.
	
	\bibitem[Bak et al. 1988]{soc} Bak, P.,  Tang, C.,   Wiesenfeld, K. 1988. Self-organized criticality. \textit{Phy. Rev. A}, \textbf{38}, 364.
	
	\bibitem[Bak et al. 1987]{soc2} Bak, P.,  Tang, C., Wiesenfeld, K. 1987 Self-organized criticality: An explanation of the 1/f noise. \textit{Physical review letters}, 59(4), 381.
	
	\bibitem[Bak 1996]{soc3}  Bak, P. 1996. How nature works: the science of self-organized criticality. \textit{Springer Science \& Business Media}.
	
	\bibitem[Bak et al 1989]{soc4}  Bak, P.,  Chen, K.,  Creutz, M. 1989. Self-organized criticality in the ``Game of Life''. \textit{Nature}, 342(6251), 780.
	
	\bibitem[Biondo et al. 2015]{finantial1}  Biondo,A. E.,  Pluchino,A.,  Rapisarda, A. 2015. Modeling financial markets by self-organized criticality. \textit{Physical Review E}, 92(4), 042814.
	
	\bibitem[Mantegna and Stanley 1995]{economy6}  Mantegna,R.N.,  Stanley,H.E. 1995. Scaling behaviour in the dynamics of an economic index. \textit{Nature}, 376(6535), 46.

	
	\bibitem[Bak et al. 1993]{economy1}  Bak, P.,  Chen, K.,  Scheinkman, J.,  Woodford, M. 1993. Aggregate fluctuations from independent sectoral shocks: self-organized criticality in a model of production and inventory dynamics. \textit{Ricerche Economiche}, 47(1), 3-30.
	
	\bibitem[Scheinkman and Woodford 1994]{economy2}  Scheinkman,J.A.,  Woodford, M. 1994. Self-organized criticality and economic fluctuations. \textit{The American Economic Review}, 84(2), 417-421.
	
	\bibitem[Andergassen et al. 2006]{economy3} Andergassen,R.,   Nardini,F.,   Ricottilli,M. 2006. Innovation waves, self-organized criticality and technological convergence. \textit{Journal of Economic Behavior \& Organization}, 61(4), 710-728.
	
	\bibitem[Batty and Xie 1999]{economy4} Batty, M.,  Xie, Y. 1999.  Self-organized criticality and urban development. \textit{Discrete Dynamics in Nature and Society}, 3(2-3), 109-124.
	
	\bibitem[Cuniberti 2001]{economy5} Cuniberti, G.,  Valleriani,A. ,   Vega, J.L. 2001. Effects of regulation on a self-organized market. \textit{Quantitative Finance}, 1(3), 332-335.
	
	
	\bibitem[Youjae 1990]{priming3}  Youjae,I. 1990. Cognitive and Affective Priming Effects of the Context for Print Advertisements,  \textit{Journal of Advertising}, \textbf{19}, 40-48.
	
	\bibitem[Tulving and Schacter 1990]{priming1} Tulving,E.,  Schacter, D. 1990. Priming and human memory systems. \textit{Science}, \textbf{247}, 301-306.
	
	\bibitem[Neely 1977]{priming2}  Neely, J. 1977. Semantic priming and retrieval from lexical memory: Roles of inhibitionless spreading activation and limited-capacity attention.  \textit{Journal of experimental psychology: general}, \textbf{106}, 226.
	
	
	
	
	\bibitem[Swart 2017]{swart} Swart J.M. 2017. A Simple Rank-Based Markov Chain with Self-Organized Criticality. \textit{Markov Processes And Related Fields}, \textbf{23},  87-102.	

	\bibitem[Van Buskirk 2007]{radiohead}  Van Buskirk,E. 2007. 2 out of 5 Downloaders Paid for Radiohead?s ``In
	Rainbows''. \textit{Wired Magazine}, 05, 47.

\bibitem[Suzdaltsev 2020]{worst_case} Suzdaltsev, A. 2020. An optimal distributionally robust auction. arXiv preprint arXiv:2006.05192.
	
	
	\bibitem[Ariely 2003]{sub0}  Ariely, D. 2003. I. Simonson, Buying, Bidding, Playing, or Competing Value Assessment and Decision Dynamics in Online Auctions. \textit{Journal of Consumer Psychology},
	\textbf{13}, 113-123.
\bibitem[Fraiman and Nistico]{marco2} Fraiman D, Nistico M (in preparation).



\end{thebibliography}

\end{document}